\documentclass[sigconf,screen]{acmart}

\AtBeginDocument{%
  \providecommand\BibTeX{{%
    \normalfont B\kern-0.5em{\scshape i\kern-0.25em b}\kern-0.8em\TeX}}}

\setcopyright{acmlicensed}
\acmPrice{15.00}
\acmDOI{10.1145/3617572.3617879}
\acmYear{2023}
\copyrightyear{2023}
\acmSubmissionID{fsews23sddmain-p28-p}
\acmISBN{979-8-4007-0377-5/23/12}
\acmConference[SDD '23]{Proceedings of the 1st International Workshop on Software Defect Datasets}{December 8, 2023}{San Francisco, CA, USA}
\acmBooktitle{Proceedings of the 1st International Workshop on Software Defect Datasets (SDD '23), December 8, 2023, San Francisco, CA, USA}
\received{2023-07-31}
\received[accepted]{2023-08-21}

\usepackage{amsmath,amsfonts}
\usepackage{algorithmic}
\usepackage{graphicx}
\usepackage{textcomp}
\usepackage{subcaption}
\usepackage{xcolor}
\usepackage{multirow}
\usepackage{paralist}
\usepackage{comment}
\usepackage{tabularx}
\newtheorem{myprob}{Problem}

\begin{document}

\title{Code Revert Prediction with Graph Neural Networks: A Case Study at J.P. Morgan Chase}


\author{Yulong Pei}
\email{yulong.pei@jpmchase.com}
\orcid{0000-0003-3739-5627}
\affiliation{%
  \institution{J.P. Morgan AI Research}
  \city{London}
  \country{UK}
}
\author{Salwa Alamir}
\email{salwa.alamir@jpmchase.com}
\orcid{0009-0006-6650-7041}
\affiliation{%
  \institution{J.P. Morgan AI Research}
  \city{London}
  \country{UK}
}
\author{Rares Dolga}
\email{rares.dolga@jpmchase.com}
\orcid{0000-0002-1800-411X}
\affiliation{%
  \institution{J.P. Morgan AI Research}
  \city{London}
  \country{UK}
}
\author{Sameena Shah}
\email{sameena.shah@jpmchase.com}
\orcid{0009-0000-5960-5811}
\affiliation{%
  \institution{J.P. Morgan AI Research}
  \state{New York}
  \country{USA}
}

\renewcommand{\shortauthors}{Yulong Pei, Salwa Alamir, Rares Dolga, and Sameena Shah}

\begin{abstract}
Code revert prediction, a specialized form of software defect detection, aims to forecast or predict the likelihood of code changes being reverted or rolled back in software development. This task is very important in practice because by identifying code changes that are more prone to being reverted, developers and project managers can proactively take measures to prevent issues, improve code quality, and optimize development processes. However, compared to code defect detection, code revert prediction has been rarely studied in previous research. Additionally, many previous methods for code defect detection relied on independent features but ignored relationships between code scripts. Moreover, new challenges are introduced due to constraints in an industry setting such as company regulation, limited features and large-scale codebase. To overcome these limitations, this paper presents a systematic empirical study for code revert prediction that integrates the code import graph with code features. Different strategies to address anomalies and data imbalance have been implemented including graph neural networks with imbalance classification and anomaly detection. We conduct the experiments on real-world code commit data within J.P. Morgan Chase 
which is extremely imbalanced in order to make a comprehensive comparison of these different approaches for the code revert prediction problem.
\end{abstract}





\maketitle

\section{Introduction}
The area of AI applied to software engineering tasks has been growing over the years, especially source code analysis, has grown over the years. Works in vulnerability analysis \cite{ghaffarian2017software}, quality assessment \cite{reddivari2019software}, testing \cite{durelli2019machine}, and code maintenance \cite{alamir2022ai} have been completed. Early defect prediction offers substantial benefits, as it reduces long-term costs \cite{shrikanth2021early}. Predicting potential production issues in code is especially advantageous in an industry setting.

Software engineering extensively explores defect detection using machine learning \cite{fenton1999critique, menzies2010defect, wei2019establishing, lessmann2008benchmarking}. These approaches utilize various features, such as code metadata, developer experience, and file-related information. Just-in-time (JIT) defect detection \cite{yang2015deep} has gained significant attention, aiming to predict bugs at the change-level \cite{kim2008classifying}. The latest JIT methods employ advanced machine learning and deep learning models, learning code representations from multiple inputs like code changes and commit messages \cite{hoang2020cc2vec,hoang2019deepjit}.

\begin{table*}
\caption{List of features used in this study.}
\label{tb:feat}
\small
\begin{tabular}{l|c|l}
\hline
\textbf{Feature}            & \textbf{Importance (IV)}             & \textbf{Relationships to code reverts}    \\\hline
Revert frequency last 30 days       &   0.570    & Reverting within the last 30 days corresponds to an increased likelihood in another revert.                                             \\\hline
File version                   &    0.326        & High file versions correspond to an increased likelihood of reverting.                                                                  \\\hline
Commit to push lag days      &    0.188   & Longer lag between commit and push is associated with a higher revert rate. \\\hline
Total lines of code in push set    &  0.151   & More lines of code leads to a higher revert rate.                                                             \\\hline
Total Cyclomatic complexity      &  0.100   & Higher total complexity corresponds to increased likelihood with reverting.                                                             \\\hline
Number of unique contributors  & 0.082 & A higher number of contributors corresponds to an increased likelihood of reverting.                                                    \\\hline
Number of dependent modules         & 0.063 & A higher number of dependencies is more likely to result in a revert.                                           \\\hline
Number of files in push set        &  0.014  & Small changes are less likely to revert.   \\\hline                                                                    
\end{tabular}
\vspace*{-0.25cm}
\end{table*}

In an industrial environment, code defect detection faces different constraints and requirements. When dealing with production issues, there are two common resolution methods: fixing forward (adding new code) or rolling back to the last working version. Rolling back is preferred for severe and time-critical issues, with fix forward implemented later. This type of commit is termed a ``risky commit''. However, industrial environments pose additional challenges for defect detection. The large-scale codebases in these companies make it impractical to analyze code line by line using previous JIT methods that utilized Abstract Syntax Tree (AST) and data flow graph (DFG). Moreover, limited access to code attributes and content hinders the use of some effective features from previous defect detection methods.

In this paper, we propose a novel problem, code revert prediction, arising from real-world industrial environments. Code revert prediction is a novel and specialized form of software defect detection. Different from traditional defect detection, it forecasts the probability of code changes being rolled back during software development, carrying significant practical significance as it allows proactive measures to prevent issues, improve code quality, and optimize development. Early prediction of code reversion effectively mitigates potential risks, especially in industrial settings where reverted issues are more critical than typical defects in production. Additionally, the problem benefits from access to historical data from code commit logs within the company, eliminating the need for data annotation or other methods to obtain labels. Despite its practical importance, research on code revert prediction remains scarce in software engineering.

\begin{figure}
    \centering
    \includegraphics[width=0.46\textwidth]{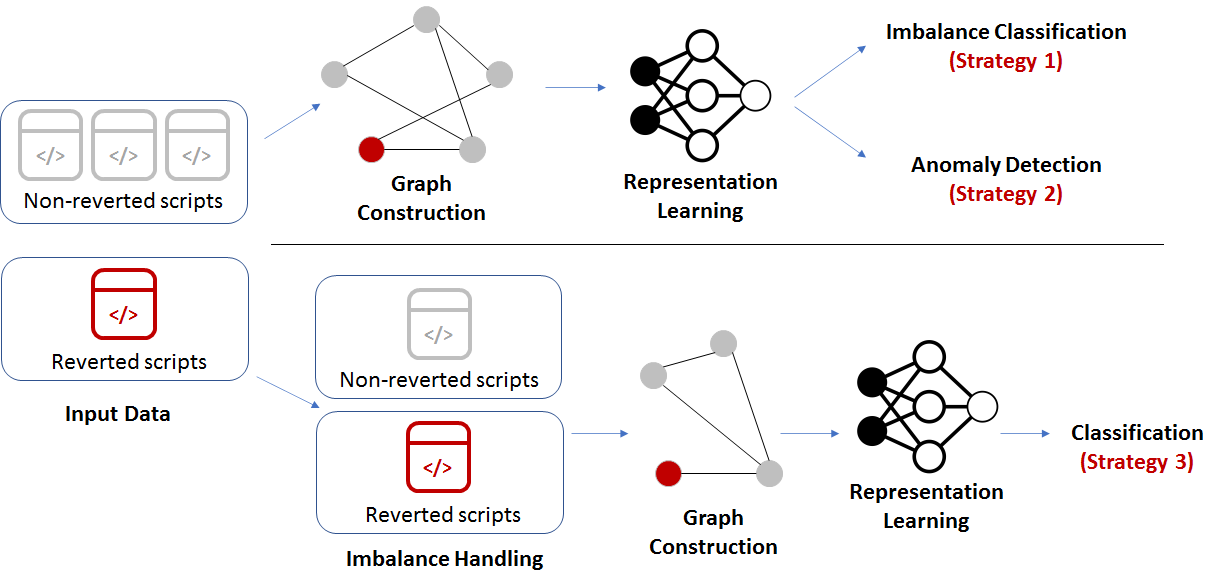}
    \caption{Different strategies to detect code reverts.}
    \label{fig:strategies}
\vspace*{-0.7cm}
\end{figure}

Code revert prediction can be formulated as a binary classification problem where the predicted labels are if the code script will be reverted or not. One can follow previous defect detection methods to construct classifiers to detect reverts. However, many previous methods that are designed for traditional defect detection tasks do not incorporate the dependencies between code which may be an important feature used for prediction. As a result, more recent works on code representation learning and defect detection have investigated the use of graphs and Graph Neural Networks (GNNs) \cite{wu2020comprehensive} e.g., \cite{allamanis2018learning,wang2020learning}. Nonetheless, these approaches rely on the abstract syntax tree of the code (AST) and/or code running logic (DFG) that are obtained via a dynamic analysis. In an industry setting (particularly a regulated one) we are constrained such that we are unable to run or even access to millions of lines of production code in order to construct this tree. Therefore, we must resort to a static analysis of the codebase in order to obtain a representation of the dependencies between the code. 

In this paper, we present a comprehensive empirical investigation into the prediction of risky code commits that are likely to be reverted. We construct a code graph using code dependencies, i.e., code import relationships. Our study specifically focuses on leveraging GNNs and introduces various strategies to address this problem. Since the distribution of code commit data is highly imbalanced, with less than 4\% of code commits resulting in reverts, we also explore two distinct approaches: anomaly detection and imbalance classification.
In summary, our contributions include: 
\begin{compactitem}
    \item We propose a novel problem formulation, i.e., code revert prediction, which is a specialized form of defect detection and aims to predict the likelihood of code changes being reverted or rolled back in software development, which is more practical in industrial settings. To the best of our knowledge, this is the first study on code revert prediction in an industrial environment.
    \item We empirically and systematically study code revert prediction problem by incorporating code dependencies, i.e., code import relationships, and using GNNs from both an anomaly detection and imbalance classification perspective.
    \item We discuss in detail promising future directions for this challenge including imbalance classification and explainability, which could be of interest to the research community.
\end{compactitem}

\section{Methodology}
\label{method}

\subsection{Problem Statement}
We first formulate the problem of code revert prediction by considering real-world constraints in industrial settings. It is intuitive that the relationships between code scripts may play vital role in identifying code importance and riskiness. Therefore, we propose to construct a code graph to capture the relationships between code. Specifically, we make use the import information\footnote{In this paper, we only study the Python code and details are shown in Section \ref{exp}. For other programming languages, similar dependency relations can be extracted to construct the code graph.}. Moreover, we ignore the direction of import relationship, so the constructed code graph is undirected. The problem is formally stated as:
\begin{myprob}
\textbf{Code revert prediction}. Consider a code graph $G=\{V,E,X\}$, where $V=\{v_1,v_2,...,v_n\}$ is the set of $n$ nodes and each node $v_i$ represents a code script, $E=\{e_{ij}\}\subseteq V\times V$ is the set of edges and each edge $e_{ij}$ represents the import relationship between code script $v_i$ and $v_j$, and $X\in \mathbb{R}^{n\times m}$ is a set of node attributes and $m$ represents the number of attributes. Assume each node is assigned to a label $y_i\in L=\{0, 1\}$ where $y_i=1$ indicates script $n_i$ is reverted and $y_i=1$ means non-reverted, and we have known the labels of a set of nodes $V_L$. The objective of code revert prediction is to predict the labels of nodes in $V\backslash V_L$.
\end{myprob}
Note that in real world, the number of commits that result in code reverts is much smaller than normal code commits that leads to the extreme imbalance in the data.

\subsection{Features}

To accurately predict code reverts involving both code and developer information, we utilize a comprehensive set of features. This set encompasses code-related and developer-related information, taking into account relationships to code reverts and features from previous studies \cite{dejaeger2012toward}. The detailed list of features and their relationships to code reverts can be found in Table~\ref{tb:feat}. We also provide the importance of features with Information Value (IV). Features with higher IV values are generally considered to be more important in predicting the target variable.

\subsection{Framework}
One major challenge in code revert prediction is the extremely imbalanced data distribution, where less than 4\% of code commits result in a revert. To address this, we consider treating the problem as graph imbalanced classification or graph anomaly detection. We employed three different strategies to solve the problem, as depicted in Fig. \ref{fig:strategies}. These strategies are outlined in detail below:
\begin{itemize}
    \item \textbf{Strategy 1}: We construct an import graph from code scripts through static analysis to address the product running issue. Then, a GNN learns the representation. Finally, we use imbalance classification, like upsampling and downsampling combined with a classifier, to predict code reverts.
    \item \textbf{Strategy 2}: Same as Strategy 1 to learn code representation, but we use anomaly detection method to identify code revert.
    \item \textbf{Strategy 3}: Given the code scripts, we first upsample (the majority) or downsample (the minority) the data to make it more balanced, then construct the import graph. Finally we use GNN to predict the revert.
\end{itemize}
In this study, we utilize advanced machine learning techniques to implement all three strategies. For effective representations, we employ both \textit{node2vec} \cite{grover2016node2vec} and Graph Convolutional Networks (GCN) \cite{kipf2016semi}. \textit{node2vec} captures information from the code dependency graph, while GCN learns representations from both graph structure and code features. To address imbalanced classes, we explore the effectiveness of upsampling, downsampling, and SMOTE \cite{smote}. Handling graph inputs is another challenge, so we employ graph imbalance learning and anomaly detection approaches, specifically designed to handle imbalanced data and identify anomalies in graph data, respectively.

\section{Experiments}
\label{exp}

\subsection{Experimental Setup}
We conduct experiments on real-world code commit data within J.P. Morgan Chase. We target the largest codebase: a Python codebase containing more than 10 million lines of code with over 3,000 code committers. We collect code commits for one month and filter out initialized and non-Python scripts. This results in about 30k commits among which less than 4\% commits were reverted. 

In Section \ref{method}, we conduct experiments covering three strategies: (1) regular classification using LR, SVM, and RF; (2) anomaly detection methods including LOF, IF, and OCSVM; and (3) imbalanced classification using Up, Down, and SMOTE. Code representations from code dependencies are learned using \textit{node2vec} (n2v) \cite{grover2016node2vec} and graph auto-encoder (replacing the supervised loss in GCN \cite{kipf2016semi} with the reconstruction loss). Additionally, we compare specially designed GNNs for anomaly detection and imbalance classification, i.e., Dominant \cite{ding2019deep} and GraphSMOTE \cite{zhao2021graphsmote}.

We use macro F1 and AUC-ROC score as the evaluation metrics to verify the performance since these are standard metrics in literature for JIT defect detection \cite{hoang2019deepjit,pornprasit2021jitline,pornprasit2022deeplinedp}. For these supervised methods, the split ration for training and test set is 80:20. To make a fair comparison, for these unsupervised methods, i.e., anomaly detection models, we only conduct experiments on the test set.

\begin{table*}
\centering
\caption{Code revert detection results using Strategy 1 and 2 w.r.t AUC-ROC score.}
\label{tb:auc_roc}
\small
\begin{tabular}{cc|c|c|clccl}
\hline
\multicolumn{2}{|l|}{}                                                                                                              & Attributes   & Structure & \multicolumn{5}{c|}{Attributes + Structure}                                                                                            \\ \hline
\multicolumn{2}{|c|}{Model}                                                                                                         & raw features & node2vec  & \multicolumn{2}{c|}{node2vec + raw features} & \multicolumn{1}{c|}{Graph Auto-Encoder (GAE)} & \multicolumn{2}{l|}{GAE + raw features} \\ \hline
\multicolumn{1}{|c|}{\multirow{3}{*}{\begin{tabular}[c]{@{}c@{}}Regular\\ Classification\end{tabular}}}    & LR    & 0.5120       & 0.5000    & \multicolumn{2}{c|}{0.5239}                  & \multicolumn{1}{c|}{0.5000}                   & \multicolumn{2}{c|}{0.5181}             \\ \cline{2-9} 
\multicolumn{1}{|c|}{}                                                                                     & SVM & 0.5000       & 0.5000    & \multicolumn{2}{c|}{0.5000}                  & \multicolumn{1}{c|}{0.5000}                   & \multicolumn{2}{c|}{0.5000}             \\ \cline{2-9} 
\multicolumn{1}{|c|}{}                                                                                     & RF          & 0.5000       & 0.5000    & \multicolumn{2}{c|}{0.5000}                  & \multicolumn{1}{c|}{0.5000}                   & \multicolumn{2}{c|}{0.4964}             \\ \hline
\multicolumn{1}{|c|}{\multirow{3}{*}{\begin{tabular}[c]{@{}c@{}}Anomaly\\ Detection\end{tabular}}}         & LOF   & 0.5780       & 0.4713    & \multicolumn{2}{c|}{0.4723}                  & \multicolumn{1}{c|}{0.4696}                   & \multicolumn{2}{c|}{0.5079}             \\ \cline{2-9} 
\multicolumn{1}{|c|}{}                                                                                     & IF       & 0.6063       & 0.4631    & \multicolumn{2}{c|}{0.4980}                  & \multicolumn{1}{c|}{0.4905}                   & \multicolumn{2}{c|}{0.4928}             \\ \cline{2-9} 
\multicolumn{1}{|c|}{}                                                                                     & OCSVM          & 0.5370       & 0.4775    & \multicolumn{2}{c|}{0.5726}                  & \multicolumn{1}{c|}{0.4951}                   & \multicolumn{2}{c|}{0.5376}             \\ \hline
\multicolumn{1}{|c|}{\multirow{3}{*}{\begin{tabular}[c]{@{}c@{}}Imbalanced\\ Classification\end{tabular}}} & Up             & 0.6969       & 0.7038    & \multicolumn{2}{c|}{0.7158}                  & \multicolumn{1}{c|}{0.4924}                   & \multicolumn{2}{c|}{0.6941}             \\ \cline{2-9} 
\multicolumn{1}{|c|}{}                                                                                     & Down           & 0.6809       & 0.6694    & \multicolumn{2}{c|}{0.7052}                  & \multicolumn{1}{c|}{0.4871}                   & \multicolumn{2}{c|}{0.6808}             \\ \cline{2-9} 
\multicolumn{1}{|c|}{}                                                                                     & SMOTE                  & 0.5156       & 0.7080    & \multicolumn{2}{c|}{\textbf{0.7228}}                  & \multicolumn{1}{c|}{0.4731}                   & \multicolumn{2}{c|}{0.6799}             \\ \hline
\end{tabular}
\vspace*{-0.25cm}
\end{table*}

\begin{table*}
\centering
\caption{Code revert detection results using Strategy 1 and 2 w.r.t Macro F1.}
\label{tb:f1}
\small
\begin{tabular}{|cc|c|c|clccl|}
\hline
\multicolumn{2}{|l|}{}                                                                                                              & Attributes   & Structure & \multicolumn{5}{c|}{Attributes + Structure}                                                                                            \\ \hline
\multicolumn{2}{|c|}{Model}                                                                                                         & raw features & node2vec  & \multicolumn{2}{c|}{node2vec + raw features} & \multicolumn{1}{c|}{Graph Auto-Encoder (GAE)} & \multicolumn{2}{l|}{GAE + raw features} \\ \hline
\multicolumn{1}{|c|}{\multirow{3}{*}{\begin{tabular}[c]{@{}c@{}}Regular\\ Classification\end{tabular}}}    & LR    & 0.5200       & 0.4964    & \multicolumn{2}{c|}{\textbf{0.5414}}                  & \multicolumn{1}{c|}{0.4964}                   & \multicolumn{2}{c|}{0.5181}             \\ \cline{2-9} 
\multicolumn{1}{|c|}{}                                                                                     & SVM & 0.4964       & 0.4964    & \multicolumn{2}{c|}{0.4964}                  & \multicolumn{1}{c|}{0.4964}                   & \multicolumn{2}{c|}{0.4964}             \\ \cline{2-9} 
\multicolumn{1}{|c|}{}                                                                                     & RF          & 0.4964       & 0.4964    & \multicolumn{2}{c|}{0.4964}                  & \multicolumn{1}{c|}{0.4964}                   & \multicolumn{2}{c|}{0.4964}             \\ \hline
\multicolumn{1}{|c|}{\multirow{3}{*}{\begin{tabular}[c]{@{}c@{}}Anomaly\\ Detection\end{tabular}}}         & LOF   & 0.4850       & 0.4769    & \multicolumn{2}{c|}{0.4699}                  & \multicolumn{1}{c|}{0.4311}                   & \multicolumn{2}{c|}{0.4287}             \\ \cline{2-9} 
\multicolumn{1}{|c|}{}                                                                                     & IF       & 0.4531       & 0.4757    & \multicolumn{2}{c|}{0.4981}                  & \multicolumn{1}{c|}{0.4916}                   & \multicolumn{2}{c|}{0.4928}             \\ \cline{2-9} 
\multicolumn{1}{|c|}{}                                                                                     & OCSVM          & 0.5128       & 0.4842    & \multicolumn{2}{c|}{0.5252}                  & \multicolumn{1}{c|}{0.4912}                   & \multicolumn{2}{c|}{0.5137}             \\ \hline
\multicolumn{1}{|c|}{\multirow{3}{*}{\begin{tabular}[c]{@{}c@{}}Imbalanced\\ Classification\end{tabular}}} & Up             & 0.4350       & 0.4363    & \multicolumn{2}{c|}{0.4565}                  & \multicolumn{1}{c|}{0.3905}                   & \multicolumn{2}{c|}{0.4373}             \\ \cline{2-9} 
\multicolumn{1}{|c|}{}                                                                                     & Down           & 0.4257       & 0.4059    & \multicolumn{2}{c|}{0.4279}                  & \multicolumn{1}{c|}{0.3861}                   & \multicolumn{2}{c|}{0.4255}             \\ \cline{2-9} 
\multicolumn{1}{|c|}{}                                                                                     & SMOTE                  & 0.5047       & 0.4352    & \multicolumn{2}{c|}{0.4580}                  & \multicolumn{1}{c|}{0.3544}                   & \multicolumn{2}{c|}{0.4343}             \\ \hline
\end{tabular}
\vspace*{-0.25cm}
\end{table*}

\subsection{Experimental Results}
Experimental results for Strategy 1 and 2 introduced in Section \ref{method} are shown in Table~\ref{tb:auc_roc} and \ref{tb:f1}, Note that for the imbalanced classification, we use LR as the classifier since it achieves the best performance compared to other traditional classifiers. From these results, some observations can be made as follows:
\begin{itemize}
    \item It becomes evident that detecting code riskiness poses a significant challenge, as indicated by the overall relatively low F1 and AUC-ROC scores across all methods. However, by combining attributes and structures, better performance can be achieved. For example, \textbf{node2vec}+raw features achieves the best performance.
    \item Traditional classifiers struggle to handle the imbalanced nature of the learning task. These models fail to identify any code reverts, highlighting their limitations in this context. Surprisingly, even complex models like random forest demonstrate poorer performance compared to simpler approaches such as logistic regression.
    \item Imbalanced learning methods outperform anomaly detection techniques. This finding suggests that defining anomalies specifically in the context of code riskiness proves to be a more intricate task. General concepts of outliers or anomalies may not effectively capture the nuanced characteristics of risky code instances.
\end{itemize}

\begin{table}
\centering
\caption{Performance comparison with Strategy 3 and GNNs.}
\label{tb:gnn}
\small
\begin{tabular}{|l|c|c|}
\hline
Model              & \multicolumn{1}{l|}{AUC-ROC} & \multicolumn{1}{l|}{Macro F1} \\ \hline
SMOTE (Strategy 1)   & 0.7228                       & 0.4580                        \\ \hline
OCSVM (Strategy 2)   & 0.5726                       & 0.5252                        \\ \hline
GCN                & 0.4964                       & 0.5000                        \\ \hline
Downsampling + GCN (Strategy 3) & \textbf{0.7269}                       & \textbf{0.5695}                        \\ \hline
GraphSMOTE \cite{zhao2021graphsmote}& 0.6423                       & 0.5176                        \\ \hline
Dominant \cite{ding2019deep} & 0.5557                       & 0.5255                        \\ \hline
\end{tabular}
\vspace*{-0.45cm}
\end{table}

We implement Strategy 3 and compare the results. Additionally, we explore the problem from the perspectives of graph anomaly detection and imbalance classification, comparing state-of-the-art GNNs for anomaly detection (Dominant \cite{ding2019deep}) and imbalance classification (GraphSMOTE \cite{zhao2021graphsmote}). Table \ref{tb:gnn} shows the results (including the best performances from Strategies 1 and 2).From these results, it can be observed that:
\begin{itemize}
    \item The best performance comes from combining downsampling and GCN, indicating the dataset's imbalance significantly impacts prediction. Downsampling + GCN consistently outperforms all other methods in both metrics, even compared to the best performers from Strategies 1 and 2.
    \item Specific GNNs for anomaly detection (Dominant) and imbalance classification (GraphSMOTE) improve performance but are still unsatisfactory and perform worse than Downsampling + GCN. 
\end{itemize}

The results emphasize the need for tailored approaches to address code revert prediction challenges. Despite additional challenges compared to JIT defect detection \cite{pornprasit2021jitline}, our performance is satisfactory. Utilizing attributes, structures, downsampling techniques, and specialized imbalance learning methods can improve code revert identification. It also calls for the development of novel techniques considering the unique nature of code reverts beyond conventional anomaly detection.

\section{THREATS TO VALIDITY}
The orientation of the solution towards production impose certain limitations on our research. We would like to highlight following important threats to validity.

\textbf{Graph Construction}. There are more fine-grained graph construction methods. Code import is unidirectional, such direction may contain important information. Moreover, other relationships such as push sets could be informative in detecting riskiness. Capturing these relationships may further improve the performance.

\textbf{Imbalance}. The extremely imbalanced distribution is the main challenge. As shown in the experiments, general graph imbalance classification and anomaly detection approaches cannot achieve promising results. Therefore, how to better handle the imbalance issue in code revert prediction in order to ultimately enhance the prediction performance, is worth to explore in the future. 

\textbf{Explainability}. Apart from achieving good performance, explaining results is crucial, especially when using black-box models like neural networks. Enhancing the interpretability for code revert prediction can provide valuable insights and foster confidence in their predictions, promoting their adoption in real-world scenarios. 

\textbf{Noisy Labels}. One commit can consist of multiple code scripts. Currently, if one of the scripts has issues and is reverted, all the committed scripts will be labeled as reverts. Such labels may bring noise to the data. Thus, finer-grained revert labels will be beneficial for this problem.

Although it is important to mention these threats, we believe that they do not invalidate the usefulness of this study and the empirical results.

\section{Related Work}
\label{related}
Defect detection in software engineering has been extensively studied, using early methods like functional and structural testing \cite{kamsties1995empirical}. Later approaches employ traditional machine learning techniques such as PCA \cite{ceylan2006software} and SVM \cite{mockus2000predicting} with features like change message terms and added and deleted line changes. An empirical comparison of these methods is presented in \cite{tantithamthavorn2016empirical}. Deep learning has also shown promise in code defect detection \cite{wang2018deep,wang2016automatically}.

Just-in-time (JIT) defect detection, a special case of defect detection, has gained attention \cite{yang2015deep}. JIT aims to identify defects at the change-level \cite{kim2008classifying}, enabling detection and fixing during development. Machine learning, particularly deep learning methods, have been applied to this problem \cite{hoang2019deepjit,pornprasit2022deeplinedp,pornprasit2021jitline}. DeepJIT uses two CNNs to detect defects in code changes and commit messages \cite{hoang2019deepjit}. DeepLineDP learns semantic properties of tokens and lines to identify defective files and lines \cite{pornprasit2022deeplinedp}. JITLine integrates traditional machine learning techniques with comparable performance \cite{pornprasit2021jitline}. Recently, to enhance code representation learning, methods have explored code relationships using AST and data flow graphs (DFG). For instance, Gated Graph Neural Networks have been applied on AST to learn program representations \cite{allamanislearning}. Devign \cite{zhou2019devign} combines Gated Graph Recurrent and convolution layers on AST and DFGs for vulnerability identification. GINN \cite{wang2020learning} generalizes graph neural networks on AST to learn semantic embeddings of source code.

Different from previous studies on code defect detection, in this paper, we explore a new task named code revert prediction and focus on a different type of code graph because of real-world constrains in industrial settings.

\section{Conclusion}
We have conducted a systematic empirical study for code riskiness prediction. Both independent code features and code import dependencies have been incorporated for the experimental studies. Graph neural networks as well as imbalance classification and anomaly detection have been compared. The experimental studies are conducted on a labeled dataset of code commit records from real-world projects within J.P. Morgan Chase . We also discussed several promising future directions to further improve the performance.

\section*{Acknowledgements}
\textbf{Disclaimer} This paper was prepared for informational purposes by the Artificial Intelligence Research group of JPMorgan Chase \& Co and its affiliates (``JP Morgan”), and is not a product of the Research Department of JP Morgan. JP Morgan makes no representation and warranty whatsoever and disclaims all liability, for the completeness, accuracy or reliability of the information contained herein. This document is not intended as investment research or investment advice, or a recommendation, offer or solicitation for the purchase or sale of any security, financial instrument, financial product or service, or to be used in any way for evaluating the merits of participating in any transaction, and shall not constitute a solicitation under any jurisdiction or to any person, if such solicitation under such jurisdiction or to such person would be unlawful.
\bibliographystyle{ACM-Reference-Format}
\bibliography{sample-base}


\begin{thebibliography}{32}


\ifx \showCODEN    \undefined \def \showCODEN     #1{\unskip}     \fi
\ifx \showDOI      \undefined \def \showDOI       #1{#1}\fi
\ifx \showISBNx    \undefined \def \showISBNx     #1{\unskip}     \fi
\ifx \showISBNxiii \undefined \def \showISBNxiii  #1{\unskip}     \fi
\ifx \showISSN     \undefined \def \showISSN      #1{\unskip}     \fi
\ifx \showLCCN     \undefined \def \showLCCN      #1{\unskip}     \fi
\ifx \shownote     \undefined \def \shownote      #1{#1}          \fi
\ifx \showarticletitle \undefined \def \showarticletitle #1{#1}   \fi
\ifx \showURL      \undefined \def \showURL       {\relax}        \fi
\providecommand\bibfield[2]{#2}
\providecommand\bibinfo[2]{#2}
\providecommand\natexlab[1]{#1}
\providecommand\showeprint[2][]{arXiv:#2}

\bibitem[Alamir et~al\mbox{.}(2022)]%
        {alamir2022ai}
\bibfield{author}{\bibinfo{person}{Salwa Alamir}, \bibinfo{person}{Petr
  Babkin}, \bibinfo{person}{Nacho Navarro}, {and} \bibinfo{person}{Sameena
  Shah}.} \bibinfo{year}{2022}\natexlab{}.
\newblock \showarticletitle{AI for Automated Code Updates}. In
  \bibinfo{booktitle}{\emph{2022 IEEE/ACM 44th International Conference on
  Software Engineering: Software Engineering in Practice (ICSE-SEIP)}}. IEEE,
  \bibinfo{pages}{25--26}.
\newblock


\bibitem[Allamanis et~al\mbox{.}(2018a)]%
        {allamanis2018learning}
\bibfield{author}{\bibinfo{person}{Miltiadis Allamanis}, \bibinfo{person}{Marc
  Brockschmidt}, {and} \bibinfo{person}{Mahmoud Khademi}.}
  \bibinfo{year}{2018}\natexlab{a}.
\newblock \showarticletitle{Learning to Represent Programs with Graphs}. In
  \bibinfo{booktitle}{\emph{International Conference on Learning
  Representations}}.
\newblock


\bibitem[Allamanis et~al\mbox{.}(2018b)]%
        {allamanislearning}
\bibfield{author}{\bibinfo{person}{Miltiadis Allamanis}, \bibinfo{person}{Marc
  Brockschmidt}, {and} \bibinfo{person}{Mahmoud Khademi}.}
  \bibinfo{year}{2018}\natexlab{b}.
\newblock \showarticletitle{Learning to Represent Programs with Graphs}. In
  \bibinfo{booktitle}{\emph{International Conference on Learning
  Representations}}.
\newblock


\bibitem[Ceylan et~al\mbox{.}(2006)]%
        {ceylan2006software}
\bibfield{author}{\bibinfo{person}{Evren Ceylan}, \bibinfo{person}{F~Onur
  Kutlubay}, {and} \bibinfo{person}{Ayse~B Bener}.}
  \bibinfo{year}{2006}\natexlab{}.
\newblock \showarticletitle{Software defect identification using machine
  learning techniques}. In \bibinfo{booktitle}{\emph{32nd EUROMICRO Conference
  on Software Engineering and Advanced Applications (EUROMICRO'06)}}. IEEE,
  \bibinfo{pages}{240--247}.
\newblock


\bibitem[Chawla et~al\mbox{.}(2002)]%
        {smote}
\bibfield{author}{\bibinfo{person}{Nitesh~V. Chawla}, \bibinfo{person}{Kevin~W.
  Bowyer}, \bibinfo{person}{Lawrence~O. Hall}, {and} \bibinfo{person}{W.~Philip
  Kegelmeyer}.} \bibinfo{year}{2002}\natexlab{}.
\newblock \showarticletitle{SMOTE: Synthetic Minority over-Sampling Technique}.
\newblock \bibinfo{journal}{\emph{J. Artif. Int. Res.}} \bibinfo{volume}{16},
  \bibinfo{number}{1} (\bibinfo{date}{jun} \bibinfo{year}{2002}),
  \bibinfo{pages}{321–357}.
\newblock
\showISSN{1076-9757}


\bibitem[Dejaeger et~al\mbox{.}(2012)]%
        {dejaeger2012toward}
\bibfield{author}{\bibinfo{person}{Karel Dejaeger}, \bibinfo{person}{Thomas
  Verbraken}, {and} \bibinfo{person}{Bart Baesens}.}
  \bibinfo{year}{2012}\natexlab{}.
\newblock \showarticletitle{Toward comprehensible software fault prediction
  models using bayesian network classifiers}.
\newblock \bibinfo{journal}{\emph{IEEE Transactions on Software Engineering}}
  \bibinfo{volume}{39}, \bibinfo{number}{2} (\bibinfo{year}{2012}),
  \bibinfo{pages}{237--257}.
\newblock


\bibitem[Ding et~al\mbox{.}(2019)]%
        {ding2019deep}
\bibfield{author}{\bibinfo{person}{Kaize Ding}, \bibinfo{person}{Jundong Li},
  \bibinfo{person}{Rohit Bhanushali}, {and} \bibinfo{person}{Huan Liu}.}
  \bibinfo{year}{2019}\natexlab{}.
\newblock \showarticletitle{Deep anomaly detection on attributed networks}. In
  \bibinfo{booktitle}{\emph{Proceedings of the 2019 SIAM International
  Conference on Data Mining}}. SIAM, \bibinfo{pages}{594--602}.
\newblock


\bibitem[Durelli et~al\mbox{.}(2019)]%
        {durelli2019machine}
\bibfield{author}{\bibinfo{person}{Vinicius~HS Durelli},
  \bibinfo{person}{Rafael~S Durelli}, \bibinfo{person}{Simone~S Borges},
  \bibinfo{person}{Andre~T Endo}, \bibinfo{person}{Marcelo~M Eler},
  \bibinfo{person}{Diego~RC Dias}, {and} \bibinfo{person}{Marcelo~P
  Guimaraes}.} \bibinfo{year}{2019}\natexlab{}.
\newblock \showarticletitle{Machine learning applied to software testing: A
  systematic mapping study}.
\newblock \bibinfo{journal}{\emph{IEEE Transactions on Reliability}}
  \bibinfo{volume}{68}, \bibinfo{number}{3} (\bibinfo{year}{2019}),
  \bibinfo{pages}{1189--1212}.
\newblock


\bibitem[Fenton and Neil(1999)]%
        {fenton1999critique}
\bibfield{author}{\bibinfo{person}{Norman~E Fenton} {and}
  \bibinfo{person}{Martin Neil}.} \bibinfo{year}{1999}\natexlab{}.
\newblock \showarticletitle{A critique of software defect prediction models}.
\newblock \bibinfo{journal}{\emph{IEEE Transactions on software engineering}}
  \bibinfo{volume}{25}, \bibinfo{number}{5} (\bibinfo{year}{1999}),
  \bibinfo{pages}{675--689}.
\newblock


\bibitem[Ghaffarian and Shahriari(2017)]%
        {ghaffarian2017software}
\bibfield{author}{\bibinfo{person}{Seyed~Mohammad Ghaffarian} {and}
  \bibinfo{person}{Hamid~Reza Shahriari}.} \bibinfo{year}{2017}\natexlab{}.
\newblock \showarticletitle{Software vulnerability analysis and discovery using
  machine-learning and data-mining techniques: A survey}.
\newblock \bibinfo{journal}{\emph{ACM Computing Surveys (CSUR)}}
  \bibinfo{volume}{50}, \bibinfo{number}{4} (\bibinfo{year}{2017}),
  \bibinfo{pages}{1--36}.
\newblock


\bibitem[Grover and Leskovec(2016)]%
        {grover2016node2vec}
\bibfield{author}{\bibinfo{person}{Aditya Grover} {and} \bibinfo{person}{Jure
  Leskovec}.} \bibinfo{year}{2016}\natexlab{}.
\newblock \showarticletitle{node2vec: Scalable feature learning for networks}.
  In \bibinfo{booktitle}{\emph{Proceedings of the 22nd ACM SIGKDD international
  conference on Knowledge discovery and data mining}}.
  \bibinfo{pages}{855--864}.
\newblock


\bibitem[Hoang et~al\mbox{.}(2019)]%
        {hoang2019deepjit}
\bibfield{author}{\bibinfo{person}{Thong Hoang}, \bibinfo{person}{Hoa~Khanh
  Dam}, \bibinfo{person}{Yasutaka Kamei}, \bibinfo{person}{David Lo}, {and}
  \bibinfo{person}{Naoyasu Ubayashi}.} \bibinfo{year}{2019}\natexlab{}.
\newblock \showarticletitle{DeepJIT: an end-to-end deep learning framework for
  just-in-time defect prediction}. In \bibinfo{booktitle}{\emph{2019 IEEE/ACM
  16th International Conference on Mining Software Repositories (MSR)}}. IEEE,
  \bibinfo{pages}{34--45}.
\newblock


\bibitem[Hoang et~al\mbox{.}(2020)]%
        {hoang2020cc2vec}
\bibfield{author}{\bibinfo{person}{Thong Hoang}, \bibinfo{person}{Hong~Jin
  Kang}, \bibinfo{person}{David Lo}, {and} \bibinfo{person}{Julia Lawall}.}
  \bibinfo{year}{2020}\natexlab{}.
\newblock \showarticletitle{CC2Vec: Distributed representations of code
  changes}. In \bibinfo{booktitle}{\emph{Proceedings of the ACM/IEEE 42nd
  International Conference on Software Engineering}}.
  \bibinfo{pages}{518--529}.
\newblock


\bibitem[Kamsties and Lott(1995)]%
        {kamsties1995empirical}
\bibfield{author}{\bibinfo{person}{Erik Kamsties} {and}
  \bibinfo{person}{Christopher~M Lott}.} \bibinfo{year}{1995}\natexlab{}.
\newblock \showarticletitle{An empirical evaluation of three defect-detection
  techniques}. In \bibinfo{booktitle}{\emph{Software Engineering—ESEC'95: 5th
  European Software Engineering Conference Sitges, Spain, September 25--28,
  1995 Proceedings 5}}. Springer, \bibinfo{pages}{362--383}.
\newblock


\bibitem[Kim et~al\mbox{.}(2008)]%
        {kim2008classifying}
\bibfield{author}{\bibinfo{person}{Sunghun Kim}, \bibinfo{person}{E~James
  Whitehead}, {and} \bibinfo{person}{Yi Zhang}.}
  \bibinfo{year}{2008}\natexlab{}.
\newblock \showarticletitle{Classifying software changes: Clean or buggy?}
\newblock \bibinfo{journal}{\emph{IEEE Transactions on software engineering}}
  \bibinfo{volume}{34}, \bibinfo{number}{2} (\bibinfo{year}{2008}),
  \bibinfo{pages}{181--196}.
\newblock


\bibitem[Kipf and Welling(2016)]%
        {kipf2016semi}
\bibfield{author}{\bibinfo{person}{Thomas~N Kipf} {and} \bibinfo{person}{Max
  Welling}.} \bibinfo{year}{2016}\natexlab{}.
\newblock \showarticletitle{Semi-supervised classification with graph
  convolutional networks}.
\newblock \bibinfo{journal}{\emph{arXiv preprint arXiv:1609.02907}}
  (\bibinfo{year}{2016}).
\newblock


\bibitem[Lessmann et~al\mbox{.}(2008)]%
        {lessmann2008benchmarking}
\bibfield{author}{\bibinfo{person}{Stefan Lessmann}, \bibinfo{person}{Bart
  Baesens}, \bibinfo{person}{Christophe Mues}, {and} \bibinfo{person}{Swantje
  Pietsch}.} \bibinfo{year}{2008}\natexlab{}.
\newblock \showarticletitle{Benchmarking classification models for software
  defect prediction: A proposed framework and novel findings}.
\newblock \bibinfo{journal}{\emph{IEEE Transactions on Software Engineering}}
  \bibinfo{volume}{34}, \bibinfo{number}{4} (\bibinfo{year}{2008}),
  \bibinfo{pages}{485--496}.
\newblock


\bibitem[Menzies et~al\mbox{.}(2010)]%
        {menzies2010defect}
\bibfield{author}{\bibinfo{person}{Tim Menzies}, \bibinfo{person}{Zach Milton},
  \bibinfo{person}{Burak Turhan}, \bibinfo{person}{Bojan Cukic},
  \bibinfo{person}{Yue Jiang}, {and} \bibinfo{person}{Ay{\c{s}}e Bener}.}
  \bibinfo{year}{2010}\natexlab{}.
\newblock \showarticletitle{Defect prediction from static code features:
  current results, limitations, new approaches}.
\newblock \bibinfo{journal}{\emph{Automated Software Engineering}}
  \bibinfo{volume}{17} (\bibinfo{year}{2010}), \bibinfo{pages}{375--407}.
\newblock


\bibitem[Mockus and Weiss(2000)]%
        {mockus2000predicting}
\bibfield{author}{\bibinfo{person}{Audris Mockus} {and}
  \bibinfo{person}{David~M Weiss}.} \bibinfo{year}{2000}\natexlab{}.
\newblock \showarticletitle{Predicting risk of software changes}.
\newblock \bibinfo{journal}{\emph{Bell Labs Technical Journal}}
  \bibinfo{volume}{5}, \bibinfo{number}{2} (\bibinfo{year}{2000}),
  \bibinfo{pages}{169--180}.
\newblock


\bibitem[Pornprasit and Tantithamthavorn(2021)]%
        {pornprasit2021jitline}
\bibfield{author}{\bibinfo{person}{Chanathip Pornprasit} {and}
  \bibinfo{person}{Chakkrit~Kla Tantithamthavorn}.}
  \bibinfo{year}{2021}\natexlab{}.
\newblock \showarticletitle{JITLine: A simpler, better, faster, finer-grained
  just-in-time defect prediction}. In \bibinfo{booktitle}{\emph{2021 IEEE/ACM
  18th International Conference on Mining Software Repositories (MSR)}}. IEEE,
  \bibinfo{pages}{369--379}.
\newblock


\bibitem[Pornprasit and Tantithamthavorn(2022)]%
        {pornprasit2022deeplinedp}
\bibfield{author}{\bibinfo{person}{Chanathip Pornprasit} {and}
  \bibinfo{person}{Chakkrit~Kla Tantithamthavorn}.}
  \bibinfo{year}{2022}\natexlab{}.
\newblock \showarticletitle{Deeplinedp: Towards a deep learning approach for
  line-level defect prediction}.
\newblock \bibinfo{journal}{\emph{IEEE Transactions on Software Engineering}}
  \bibinfo{volume}{49}, \bibinfo{number}{1} (\bibinfo{year}{2022}),
  \bibinfo{pages}{84--98}.
\newblock


\bibitem[Reddivari and Raman(2019)]%
        {reddivari2019software}
\bibfield{author}{\bibinfo{person}{Sandeep Reddivari} {and}
  \bibinfo{person}{Jayalakshmi Raman}.} \bibinfo{year}{2019}\natexlab{}.
\newblock \showarticletitle{Software quality prediction: an investigation based
  on machine learning}. In \bibinfo{booktitle}{\emph{2019 IEEE 20th
  International Conference on Information Reuse and Integration for Data
  Science (IRI)}}. IEEE, \bibinfo{pages}{115--122}.
\newblock


\bibitem[Shrikanth et~al\mbox{.}(2021)]%
        {shrikanth2021early}
\bibfield{author}{\bibinfo{person}{NC Shrikanth}, \bibinfo{person}{Suvodeep
  Majumder}, {and} \bibinfo{person}{Tim Menzies}.}
  \bibinfo{year}{2021}\natexlab{}.
\newblock \showarticletitle{Early life cycle software defect prediction. why?
  how?}. In \bibinfo{booktitle}{\emph{2021 IEEE/ACM 43rd International
  Conference on Software Engineering (ICSE)}}. IEEE, \bibinfo{pages}{448--459}.
\newblock


\bibitem[Tantithamthavorn et~al\mbox{.}(2016)]%
        {tantithamthavorn2016empirical}
\bibfield{author}{\bibinfo{person}{Chakkrit Tantithamthavorn},
  \bibinfo{person}{Shane McIntosh}, \bibinfo{person}{Ahmed~E Hassan}, {and}
  \bibinfo{person}{Kenichi Matsumoto}.} \bibinfo{year}{2016}\natexlab{}.
\newblock \showarticletitle{An empirical comparison of model validation
  techniques for defect prediction models}.
\newblock \bibinfo{journal}{\emph{IEEE Transactions on Software Engineering}}
  \bibinfo{volume}{43}, \bibinfo{number}{1} (\bibinfo{year}{2016}),
  \bibinfo{pages}{1--18}.
\newblock


\bibitem[Wang et~al\mbox{.}(2018)]%
        {wang2018deep}
\bibfield{author}{\bibinfo{person}{Song Wang}, \bibinfo{person}{Taiyue Liu},
  \bibinfo{person}{Jaechang Nam}, {and} \bibinfo{person}{Lin Tan}.}
  \bibinfo{year}{2018}\natexlab{}.
\newblock \showarticletitle{Deep semantic feature learning for software defect
  prediction}.
\newblock \bibinfo{journal}{\emph{IEEE Transactions on Software Engineering}}
  \bibinfo{volume}{46}, \bibinfo{number}{12} (\bibinfo{year}{2018}),
  \bibinfo{pages}{1267--1293}.
\newblock


\bibitem[Wang et~al\mbox{.}(2016)]%
        {wang2016automatically}
\bibfield{author}{\bibinfo{person}{Song Wang}, \bibinfo{person}{Taiyue Liu},
  {and} \bibinfo{person}{Lin Tan}.} \bibinfo{year}{2016}\natexlab{}.
\newblock \showarticletitle{Automatically learning semantic features for defect
  prediction}. In \bibinfo{booktitle}{\emph{Proceedings of the 38th
  International Conference on Software Engineering}}.
  \bibinfo{pages}{297--308}.
\newblock


\bibitem[Wang et~al\mbox{.}(2020)]%
        {wang2020learning}
\bibfield{author}{\bibinfo{person}{Yu Wang}, \bibinfo{person}{Ke Wang},
  \bibinfo{person}{Fengjuan Gao}, {and} \bibinfo{person}{Linzhang Wang}.}
  \bibinfo{year}{2020}\natexlab{}.
\newblock \showarticletitle{Learning semantic program embeddings with graph
  interval neural network}.
\newblock \bibinfo{journal}{\emph{Proceedings of the ACM on Programming
  Languages}} \bibinfo{volume}{4}, \bibinfo{number}{OOPSLA}
  (\bibinfo{year}{2020}), \bibinfo{pages}{1--27}.
\newblock


\bibitem[Wei et~al\mbox{.}(2019)]%
        {wei2019establishing}
\bibfield{author}{\bibinfo{person}{Hua Wei}, \bibinfo{person}{Changzhen Hu},
  \bibinfo{person}{Shiyou Chen}, \bibinfo{person}{Yuan Xue}, {and}
  \bibinfo{person}{Quanxin Zhang}.} \bibinfo{year}{2019}\natexlab{}.
\newblock \showarticletitle{Establishing a software defect prediction model via
  effective dimension reduction}.
\newblock \bibinfo{journal}{\emph{Information Sciences}}  \bibinfo{volume}{477}
  (\bibinfo{year}{2019}), \bibinfo{pages}{399--409}.
\newblock


\bibitem[Wu et~al\mbox{.}(2020)]%
        {wu2020comprehensive}
\bibfield{author}{\bibinfo{person}{Zonghan Wu}, \bibinfo{person}{Shirui Pan},
  \bibinfo{person}{Fengwen Chen}, \bibinfo{person}{Guodong Long},
  \bibinfo{person}{Chengqi Zhang}, {and} \bibinfo{person}{S~Yu Philip}.}
  \bibinfo{year}{2020}\natexlab{}.
\newblock \showarticletitle{A comprehensive survey on graph neural networks}.
\newblock \bibinfo{journal}{\emph{IEEE transactions on neural networks and
  learning systems}} \bibinfo{volume}{32}, \bibinfo{number}{1}
  (\bibinfo{year}{2020}), \bibinfo{pages}{4--24}.
\newblock


\bibitem[Yang et~al\mbox{.}(2015)]%
        {yang2015deep}
\bibfield{author}{\bibinfo{person}{Xinli Yang}, \bibinfo{person}{David Lo},
  \bibinfo{person}{Xin Xia}, \bibinfo{person}{Yun Zhang}, {and}
  \bibinfo{person}{Jianling Sun}.} \bibinfo{year}{2015}\natexlab{}.
\newblock \showarticletitle{Deep learning for just-in-time defect prediction}.
  In \bibinfo{booktitle}{\emph{2015 IEEE International Conference on Software
  Quality, Reliability and Security}}. IEEE, \bibinfo{pages}{17--26}.
\newblock


\bibitem[Zhao et~al\mbox{.}(2021)]%
        {zhao2021graphsmote}
\bibfield{author}{\bibinfo{person}{Tianxiang Zhao}, \bibinfo{person}{Xiang
  Zhang}, {and} \bibinfo{person}{Suhang Wang}.}
  \bibinfo{year}{2021}\natexlab{}.
\newblock \showarticletitle{Graphsmote: Imbalanced node classification on
  graphs with graph neural networks}. In \bibinfo{booktitle}{\emph{Proceedings
  of the 14th ACM international conference on web search and data mining}}.
  \bibinfo{pages}{833--841}.
\newblock


\bibitem[Zhou et~al\mbox{.}(2019)]%
        {zhou2019devign}
\bibfield{author}{\bibinfo{person}{Yaqin Zhou}, \bibinfo{person}{Shangqing
  Liu}, \bibinfo{person}{Jingkai Siow}, \bibinfo{person}{Xiaoning Du}, {and}
  \bibinfo{person}{Yang Liu}.} \bibinfo{year}{2019}\natexlab{}.
\newblock \showarticletitle{Devign: Effective vulnerability identification by
  learning comprehensive program semantics via graph neural networks}.
\newblock \bibinfo{journal}{\emph{Advances in neural information processing
  systems}}  \bibinfo{volume}{32} (\bibinfo{year}{2019}).
\newblock


\end{thebibliography}

\end{document}